\documentclass[epjST]{svjour}
\usepackage{graphics}
\usepackage[francais, english]{babel}
\usepackage{amsmath}
\usepackage{graphicx}%
\usepackage{multirow}%
\usepackage{amsmath,amssymb,amsfonts}%
\usepackage{mathrsfs}%
\usepackage[title]{appendix}%
\usepackage{xcolor}%
\usepackage{textcomp}%
\usepackage{manyfoot}%
\usepackage{booktabs}%
\usepackage{algorithm}%
\usepackage{algorithmicx}%
\usepackage{algpseudocode}%
\usepackage{listings}%
\usepackage[numbers]{natbib}

\newcommand{\vv}{\boldsymbol{v}}
\newcommand{\id}{I}

\renewcommand{\Re}{\text{Re}}
\newcommand{\Ca}{\text{Ca}}
\newcommand{\La}{\text{La}}

\begin{document}
%
\title{Super-fast bullet bubbles transported in a pressure-driven cylindrical flow}
\author{Jean Cappello \inst{1,} \inst{2}\fnmsep\thanks{\email{jean.cappello@ulb.be}} \and Javier Rivero-Rodriguez\inst{3} \and Benoit Scheid\inst{1}}
\institute{Transfers, Interfaces and Processes (TIPs), Université Libre de Bruxelles, 1050 Brussels, Belgium \and University of Lyon, Université Claude Bernard Lyon 1, CNRS, Institut Lumière Matière,
F-69622 Villeurbanne, France. \and Escuela de Ingenierías Industriales, Universidad de Málaga, Campus de Teatinos, 29071 Málaga, Spain}
%
%



\abstract{When transported by a pressure driven flow in a cylindrical pipe, bubbles may exhibit very fast velocities. In this paper, we show that, when the bubbles are largely deformable, that is, at large capillary numbers \Ca, the velocity of the bubble can be larger than the maximal velocity of the flow that transports them. We call this regime ``super-fast''. However, the situation changes when inertia comes at play for increasing Reynolds numbers \Re, and the relative velocity of the bubble drops for sufficiently large Laplace number, defined as $\La = \Re/\Ca$. In this article, we uncover the conditions for which the super-fast regime exists : the deformability of the drop is crucial, and hence the capillary number needs to be larger than a critical value, yet smaller than a threshold above which the bubble breaks up. The two limiting capillary numbers are presented in a phase diagram as a function of the bubble size and the Laplace number.}
\maketitle

\section{Introduction}\label{sec1}

Bubbles are the key elements in most multiphase reactors that involve liquid-gas transfer. Traditionally realized in bubble columns or stirred tank, bubble flows in capillaries have gained increasing interest \cite{Zhao} not only for mass transfer enhancement \cite{Vandu} but also for flow segmentation, allowing a drastic narrowing of the residence time distribution \cite{Rojahn}. 


For these reasons, understanding the dynamics of bubbles in cylindrical capillaries at low Reynolds numbers has become crucial. The peculiar situation of long bubbles has been at the core of several seminal studies \cite{Bretherton, taylor1961deposition, cox1964experimental, Aussillous, Martinez_Udell1} that, among other things, determined the thickness of the liquid film between the bubble and the continuous phase but also the longitudinal velocity of the bubble \cite{Lac} as a function of the capillary number $\Ca = \mu J /\gamma$, where $J$ is the mean flow velocity, $\mu$ the viscosity of the liquid and $\gamma$ its surface tension. More recently, the description has been extended to the case of droplets and adjusted for couples of liquids of different viscosity ratios $\lambda = \mu_{d}/\mu$, with $\mu_d$ the viscosity of the droplets \cite{balestra2018viscous}. \newline
When the dispersed object has a size $d_\text{d}$, comparable to or smaller than the channel diameter $d_\text{h}$, the coupling between the object and the flow becomes more intricate affecting the shape and velocity of the object \cite{guido2010droplet}. 
Several experimental \cite{Olbricht, ho1975creeping, dapolito2019confined} and numerical \cite{Martinez_Udell2, Lac, dupont2007motion} studies focused on this situation. These studies particularly investigated the consequence of varying the deformability (i.e. capillary number $\Ca$) on the droplet or bubble shape and velocity. They concluded that, when the capillary number is increased, the droplet or bubble elongates, adopting a bullet-like shape while remaining centered in the channel. As a consequence, the bubble or droplet surface is farther away from the channel walls, therefore leading to an increase of its velocity. Additionally, for $\lambda \lesssim 1$ and above a threshold capillary number -- which depends on the drop or bubble size and the viscosity ratio -- an indentation (i.e., an inversion of the curvature) occurs at the rear of the bubble/droplet \cite{Olbricht, olbrichtRevue, Martinez_Udell2, Nath, Lac, cherukumudi2015prediction, Giavedoni, Goldsmith}. Another remarkable result of these studies lies in the existence of a regime, attained at large $\Ca$ for $\lambda <1/2$, where bubbles or droplets may achieve a velocity that exceeds the maximal axial velocity of the continuous phase \cite{Lac}, that is, the bubble or droplet is faster than the maximal velocity of the fluid that pushes it. This regime is hereafter denoted as the ``super-fast'' regime. Furthermore, the study of  Lac \& Sherwood \cite{Lac} shows that while for moderate capillary numbers ($\Ca <0.4$) the velocity of low viscosity drops decreases with increasing drop size, the opposite is observed for large capillary numbers. For $\lambda\lesssim1$, increasing the number of capillaries further leads to the growth of the rear indentation until it forms a reentrant jet that eventually leads to breakup \cite{Olbricht,Lac,Nath}, while for $\lambda \gtrsim1$ oscillatory motion of the interface is observed at large $\Ca$ \cite{Lac}. \newline
All of the aforementioned studies were conducted in situations where inertia could be neglected. Hence, the dispersed objects were uniquely subject to deformation-induced migration forces. In most situations, this force points toward the center line of the channel and thus stabilizes the centered position of the bubble or droplet \cite{cappello2023beads, herrada2023global, coulliette1998motion, chan1979motion}. Exceptions occur for small droplets ($d_\text{d}<0.5d_\text{h}$) of the viscosity ratio verifying $0.7\lesssim\lambda\lesssim 10$. In that situation, the droplet experiences a deformation-induced migration force oriented towards the channel wall \cite{cappello2023beads}. Furthermore, a rigid particle (i.e., small $\Ca$) transported by an inertial flow at intermediate Reynolds numbers, $\Re = \rho J d_\text{h}/\mu$ with $\rho$ the density of the continuous phase, migrates to an off-center lateral position \cite{segre1962behaviour,matas2004inertial}. The distance between the center line of the channel and this equilibrium position depends on the size of the bead. Small particles (i.e., $d_\text{d} \ll d_\text{h}$) have been shown to migrate toward the Segr\'e and Silberberg annulus of radius $0.3 d_\text{h}$ \cite{segre1962behaviour}, while beads of diameter larger than $0.8d_\text{h}$ remain centered \cite{rivero2018bubble}. In between, eccentricity decreases monotonically following a pitch-fork bifurcation \cite{cappello2023beads}.
The equilibrium position of a deformable particle, such as a bubble or a droplet, is therefore the result of a competition between the deformation-induced migration force and the inertial migration force \cite{chen2014inertia, rivero2018bubble, cappello2023beads}. It is often convenient to use the Laplace number $\La = \Re/\Ca$, to compare these two forces. At small $\La$, the deformation-induced migration force dominates and the particle is centered, while, at large $\La$, the inertial migration dominates and the particles may migrate to positions off center depending on their size. Most studies investigating the dynamics of deformable bubbles or droplets in intermediate $\Re$ and large $\Ca$ focused on long objects since they are always centered \cite{Langewisch, han2009measurement, rocha2017wide, andredaki2021, kreutzer2005inertial, heil2001finite, giavedoni1997axisymmetric}. The case of droplets or bubbles of a size comparable to the channel diameter remains less known \cite{atasi2018, chen2014inertia, rivero2018bubble, cappello2023beads}. In a previous study, we gave a formula for the derivation of the velocity of droplets or bubbles located at different lateral positions within a microchannel \cite{cappello2023beads}. However, the formula is valid only for small deformations and is therefore not applicable at large $\Ca$. Atasi et al. \cite{atasi2018} and Rivero \& Scheid \cite{rivero2018bubble} studied numerically the transport of bubbles at large capillary numbers and non-negligible inertia with Reynolds numbers of $\Re \sim 1$ and $\Re \sim 10$, respectively. Both studies show that a super-fast regime could be observed even in the presence of inertia for $\lambda \ll $ 1, $\Ca \geq 0.5$, and bubble dimensionless size $d_\text{d}/d_\text{h} \gtrsim 0.5$. However, to our knowledge, there is no trace of experimental proof of the existence of this regime at low or intermediate Reynolds numbers and no systematic exploration of the role of the Reynolds number on the bubble velocity at large $\Ca$.

In this article, we investigate the longitudinal velocity and shape of deformable bubbles ($\lambda \ll 1$) that are centered in the microchannel and transported by a pressure-driven flow at a varying Reynolds number. By combining experiments and numerical simulation, we rationalize the dependence of the bubble longitudinal velocity with its deformability, the flow inertia, and bubble sizes. By this means, we offer the first experimental proof of the existence of the super-fast regime which may be attained above a critical capillary number but below a critical Laplace number. We developed an analytical model describing the long bubble's velocity at large capillary numbers and nonzero Reynolds numbers to explain the origin of this regime and the impact of the flow inertia. Moreover, combining our results on the centered bubble velocities with the criteria for bubble break-up and for lateral migration, we determine the range of parameters for which the super-fast regime can be observed. 

\section{Experimental set-up and methods}\label{Mat&met}

The experiments consist of characterizing the stationary dynamics of dispersed bubbles transported by an external flow in a cylindrical microchannel. A schematic of the experimental setup is provided in Fig.~\ref{fig:exp}(a).

\subsection{Experimental apparatus}

\begin{figure}[t!]
 \centering
   \includegraphics[width=1\linewidth]{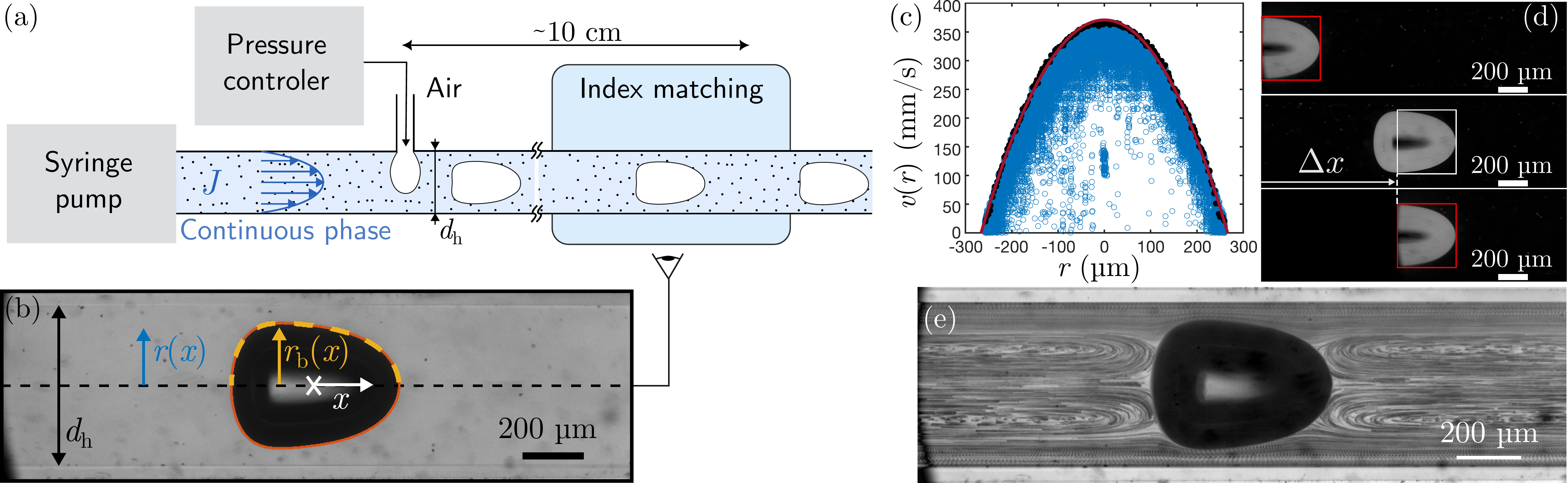}
 \caption{\textbf{Experimental setup and procedures.} \textbf{(a)} Bubbles are generated in a cylindrical capillary of diameter $d_\text{h}$ using a T-junction. The flow rate of the continuous phase is imposed using a syringe pump while the flow of air is set using a pressure controller. The cylindrical capillary is placed in a pool of mineral oil to allow visualisation. \textbf{(b)} Typical picture of a transported bubble. The shape of the bubble (red line) can be obtained through image analysis. From the bubble shape, we recover the volume of the bubble by integrating the bubble cross-section of radius $r_\text{b}(z)$ over the bubble length. \textbf{(c)} Typical measurement of the velocity of the tracers far from the bubbles as a function of their positions in the microchannel. Black dots correspond to the maximal velocities measured for each positions. The red solid line is a parabolic fit of the black dots and corresponds to the velocity profile of the flow in-between the bubbles. \textbf{(d)} Bubbles are detected (red rectangle) in all the images in which at least a portion of the bubble may be observed (top panel). Using image correlation technique we look for the same portion of bubble in a reference image where the bubble is centred (white rectangle in the middle panel). The longitudinal distance between the two rectangles $\Delta x$ is measured and the raw image is displaced by this quantity to centre the bubble (bottom panel). 
 \textbf{(e)} Streamlines in the reference frame of the bubble are obtained by superimposing all the centred images of bubbles.}
 \label{fig:exp}
 \end{figure}

As shown in Fig.~\ref{fig:exp}(a), bubbles are generated using a T-junction. Continuous phase flow is imposed using a high-precision syringe pump (Nemesys, Cetoni) with a 500 µL syringe, while the air flow is controlled using a pressure controller (MCF-EX, Fluigent). Tracers of 4.5 µm diameter (Cospheric, SIO2MS-AG-3.5 4.5M) are suspended in the continuous phase and are used for particle tracking velocimetry (PTV). After bubble generation, a two-phase flow takes place in a glass microcapillary (Postnova) of diameter $d_\text{h}$ = 527 $\pm$ 5 µm. 

The dynamics of the bubble is monitored using an inverted microscope (Nikon Eclipse-Ti) equipped with a 10x TU Plan Fluor objective (NA = 0.3) and images are recorded using a high-speed camera (Photron, NOVAs12) at a frame rate ranging from 1k to 25k fps. The observation zone is fixed and located a tenth of centimeters downstream the bubble generation in order to have a developed flow and to ensure that the bubbles have reached their equilibrium shape and velocity.
To allow precise observations of the dispersed micro-objects in the cylindrical microchannel, we limit image deformations caused by optical refraction effects by placing the microcapillary in a glass tank filled with light mineral oil (Sigma Aldrich), whose refractive index ($n_\text{oil} = 1.467$) is very close to that of glass ($n_\text{glass} = 1.470$). A typical observation is shown in Fig.~\ref{fig:exp}(b).

\subsection{Continuous phase ad dispersed phase}

In order to vary the capillary, Reynolds and Laplace numbers, we selected different liquids and controlled the flow rate of the continuous phase. We used three Newtonian fluids of different viscosity and surface tension: silicone oil V5 ($\mu = 5$ mPa, $\gamma = 19.7$ mN/m), mineral oil light ($\mu = 23.7$ mPa, $\gamma = 29.6$ mN/m), and silicone oil V500 ($\mu = 500$ mPa, $\gamma = 21.1$ mN/m). By varying the flow rate of the continuous phase from 50 µL/min to 2500 µL/min we varied the capillary and Reynolds numbers of the flow. The bubble sizes were adjusted by varying the imposed pressure of the air reservoir with the pressure controller (see Fig.~\ref{fig:exp}(a))).

\subsection{Image analysis and streamlines reconstruction}

We determine the bubble velocity from the raw images by tracking the position of the bubble front tip over time and calculating the slope of the resulting curve. This process is performed for at least 500 bubbles to obtain a statistically averaged velocity.

The bubble shapes are derived from the raw images by binarizing the picture to obtain the bubble contour (red line in Fig.~\ref{fig:exp}(b)). By this mean, the dependence of the bubble radius $r_\text{b}(x)$ on the axial position $x$ is derived (yellow dashed line in Fig.~\ref{fig:exp}(b)). Taking advantage of the central position of the bubble in the capillary, we derive its volume $\mathcal{V}$, by integrating the quantity $\pi r_\text{b}(x)^2$ over the axial position. In the following, we quantify the volume of the bubble by the equivalent diameter, $d_\text{d}$, defined as $d_\text{d} = \sqrt[3]{6 \mathcal{V} / \pi }$. The confinement -- or dimensionless drop size -- is then defined as the ratio $d = d_\text{d}/d_\text{h}$.\\

The superficial velocity of the biphasic flow, $J$, is obtained by particle tracking velocimetry techniques using an ImageJ plugin (MTrack 3) and a MatLab routine. Briefly, considering only the images where bubbles are absent, we track the velocity, in the frame of reference of the laboratory, of all the tracers passing through the capillary. As the depth of field of the objective is of the order of 10 µm, tracers that are not exactly at the central plane of the channel are also observed, and their velocity is measured. To ensure a measurement of the velocity exactly at the central plane of the channel, the velocities of all the tracers are reported as a function of their radial location in the microchannel, as illustrated in Fig.~\ref{fig:exp}(c) with the blue open dots. The upper contour of the cloud of points (black dots in Fig.~\ref{fig:exp}(c)) corresponds to the velocity profile in the central plane of the channel. This contour is well fitted with a parabolic profile (red line in Fig.~\ref{fig:exp}(c)), which indicates a well-developed flow. The superficial velocity, $J$, corresponding to the average velocity of the flow, is then obtained by averaging the velocity profile over the diameter of the capillary.\\

The tracers are also used to obtain the streamlines around the flowing bubbles experimentally. Contrary to the derivation of the superficial velocity, the streamlines are derived in the frame of reference of the moving bubble. For each image where at least a portion of the bubble is visible, the background is removed (see Fig.~\ref{fig:exp}(d) top), and the image is compared to a reference picture of a centered bubble (see Fig.~\ref{fig:exp}(d) middle). Using an image correlation technique, we search for the position of the visible portion of bubble (red frame in Fig.~\ref{fig:exp}(d) top) on the reference image (white frame in Fig.~\ref{fig:exp}(d) middle) and we derive the displacement distance, $\Delta x$, by which the initial image must be translated to center the bubble (see Fig.~\ref{fig:exp}(d) bottom).
Once all bubbles are centered, streamlines around the moving object are obtained by superimposing all the images (typically 5000 images). A typical result, where the background has been added back, is shown on Fig.~\ref{fig:exp}(e).

\section{Numerical method}

The dynamics of the train of bubbles is studied not only experimentally but also by means of numerical simulations. The mathematical model is provided in all details in \cite{rivero2018bubble} and is briefly recalled here for the sake of completeness.

The pressure and velocity field, $p$ and $\vv$, around the bubble are governed by the steady incompressible Navier-Stokes equations. For a liquid of density $\rho$ and viscosity $\mu$, it writes
\begin{align}
\label{eq0}
\nabla \cdot \vv = 0 \,, \quad \rho \vv \cdot \nabla \vv = \nabla \cdot \tau \,,
\end{align}
where $\tau=-p \id + \mu \left[ \nabla \vv + \left( \nabla \vv \right)^T \right]$ is the stress tensor. Analogous to the fact that the experimental image reference is taken such that the bubble is centred in the image, in the numerical simulations the system of reference is considered to be attached to the deformable bubble, the central position of which is ensured by 
\begin{align}
\int_{\mathcal{V}_B} x \,{\rm d}\mathcal{V} = 0\,,
\end{align}
where $\mathcal{V}_B$ is the domain occupied by the bubble and $x$ is the axial coordinate. In so doing, the velocity of the wall fulfils
\begin{align}
\vv(r=d_\text{h}/2) = - V_\text{b} \boldsymbol{e}_x \,,
\end{align}
where $V_\text{b}$ is the velocity of the bubble and $\boldsymbol{e}_x$ is the axial vector pointing downstream. 
At the bubble surface, the stress balance and the kinematic condition apply:
\begin{align}
\label{eq:stress_and_impermeability}
\tau \cdot \boldsymbol{n} = -p_G \boldsymbol{n} + \gamma \boldsymbol{n}  \nabla_s \cdot \boldsymbol{n} \,, \quad \vv \cdot \boldsymbol{n}  = 0 \,,
\end{align}
where $\gamma$ is the surface tension, $\boldsymbol{n} $ the outer normal vector and $p_G$ the gas pressure. The continuous and dispersed phases having impermeable boundaries, as shown in Eq.~(\ref{eq:stress_and_impermeability}), it requires to impose a pressure reference at one point for each phase. One pressure represents the absolute pressure reference that is set to 0, at one point arbitrarily chosen in the continuous phase, while the other one remains to be determined, that is, $p_G$ = $p_\text{ref}$, at one point arbitrarily chosen in the bubble. Note that $p_\text{ref}$ is an integration constant that is determined through a constraint on the bubble volume,
\begin{align}
\int_{\mathcal{V}_B} \,{\rm d}\mathcal{V} = \frac16 \pi d_\text{d}^3 \,.
\end{align}
To model the flow around a train of bubbles, we use periodic boundary conditions together with a pressure drop $\Delta p$ along each period.
\begin{align}
\vv \vert_{\boldsymbol{x}} = \vv \vert_{\boldsymbol{x}+L \boldsymbol{e}_x} 
\,, \quad
p \vert_{\boldsymbol{x}} = p \vert_{\boldsymbol{x}+L \boldsymbol{e}_x} + \Delta p \,,
\end{align}
where $L$ is the period, namely the longitudinal distance between adjacent bubbles. The pressure drop is given by the flow rate
\begin{align}
\label{eqlast}
\int_{\Sigma_\text{cross}} \left( \vv \cdot \boldsymbol{e}_x +V_\text{b}- J \right) \,{\rm d}\Sigma = 0 \,,
\end{align}
where $\Sigma_\text{cross}$ is any cross-section. 

The system of equations \eqref{eq0}-\eqref{eqlast} is adimentionalized by using the diameter of the channel $d_h$ as characteristic length, the mean flow rate $J$ as characteristic velocity, and the viscous stress $\mu J / d_h$ as characteristic stress. Its dimensionless form is then solved by means of the finite element method with Lagrangian linear elements for the pressure and quadratic for the velocity field.
The geometry of the bubble is axisymmetric and its deformation is taken into account with the help of the arbitrary Lagrangian-Eulerian (ALE) method for which the displacement of the bubble surface is imposed, following \cite{rivero2021alternative}. 

\section{Results and discussion}



\subsection{Shape and streamlines}

\begin{figure}[t!]
 \centering
   \includegraphics[width=1\linewidth]{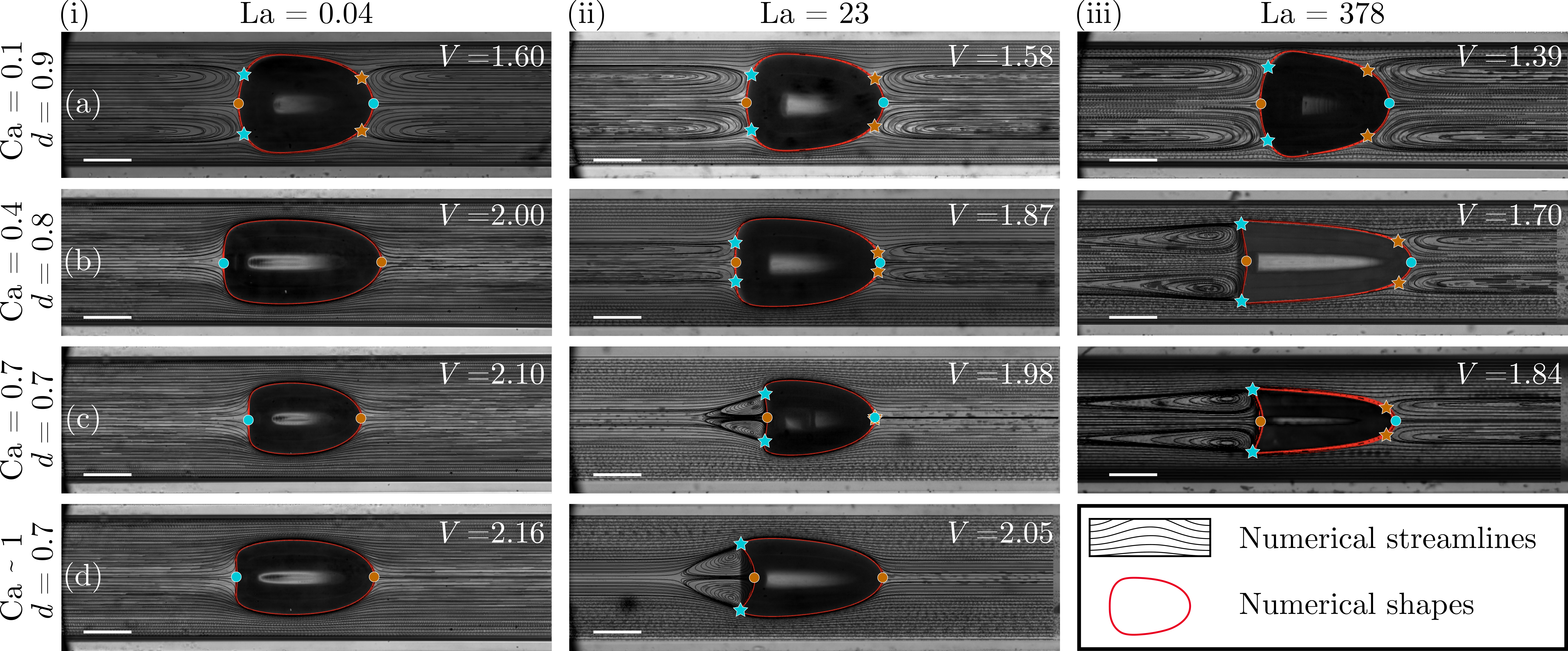}%
 \caption{\textbf{Bubble velocity and shape.} The experimentally obtained bubble shape and streamlines are compared to numerical simulations (bubble contour : red line; streamlines : black lines), for varying values of capillary number (Ca), Laplace number ($\La$) and bubble size ($d=d_\text{d}/d_\text{h}$). Circles correspond to stagnation points and stars to stagnation annuli at the bubble interface. Divergent (resp. convergent) stagnation points/annuli are coloured in blue (resp. red). From line (a) to (d) the capillary number is increased, and from column (i) to (iii) the Laplace number is increased. On each line the bubble equivalent diameter is kept constant.}
 \label{bubble_shape}
 \end{figure}
 
The shape of the bubbles and the streamlines around them are reported in Fig.~\ref{bubble_shape} for various conditions. We essentially varied the capillary number (from (a) to (d), $\Ca$ increases from 0.1 to 1), and the Laplace number (from (i) to (iii), $\La$ increases from 0.04 to 378). Numerically derived shapes (red lines) and streamlines (black lines) are superimposed on the experimental observations. The excellent agreement observed between experiments and numerical simulations validates our numerical approach and shows that the experimental method and especially the use of tracers is reasonably noninvasive and enables accurate observations.

For the smallest value of the capillary number explored here ($\Ca = 0.1$), and for $\La = 0.04$ (i.e. negligible inertia) and $\La = 23$ (Fig.~\ref{bubble_shape} line (a) columns (i) and (ii) respectively), the bubbles shapes are similar and inertia can be ignored. Similarly to what was observed by \cite{martinez1990axisymmetric} and \cite{khodaparast2015dynamics}, the deformed bubbles show a larger curvature at the front than at the back of the bubble, and, close to the channel lateral walls, the interface flattens. 
Keeping $\Ca=0.1$ while increasing the value of $\La$ to 378 (Fig.~\ref{bubble_shape}(a)(iii)) we observe that the curvature of the nose remains larger than at the back, yet the overall shape of the bubble gets quite different : inertia is no longer negligible and the bubble adopts a bell shape \cite{khodaparast2015dynamics}. 
As visible from the streamlines shown in Fig.~\ref{bubble_shape}(a), for all three values of $\La$, recirculations are present both at the front and at the back of the bubbles, meaning that the bubble is slower than the maximal flow velocity. Moreover, two stagnation annuli (stars) and two stagnation points (circles) are observed. At the bubble front, there is a divergent stagnation point on the bubble's axis of symmetry (blue circle) and a convergent stagnation annulus (orange stars), while at the bubble back, the stagnation point on the bubble axis is convergent (orange circle), and the stagnation circle is divergent (blue stars). A convergent stagnation point/annulus corresponds to a locally pushing normal stress, whereas a divergent stagnation point/annulus corresponds to a locally pulling normal stress \cite{atasi2018influence}. Hence, the inhomogeneous distribution of the normal stress along the bubble interface, generated by the recirculating flow, causes a decrease of the interface curvature near the convergent points/annuli and an increase of the curvature near the divergent points/annuli, leading to the peculiar shape of the observed bubbles. One can note that the stagnation annuli have a larger radius at the back than at the front of the bubble. Moreover, while the flow around the bubbles is similar for $\La = 0.04$ and $\La = 23$, we observe an increase of the stagnation annular radius for $\La = 378$, which is a signature of the decrease in the relative velocity of the bubble $V = V_\text{b}/J$.

When the capillary number increases, the situation changes completely. For $\Ca= 0.4$ (Fig.~\ref{bubble_shape} line (b)), the bubble becomes more deformed and elongated, adopting a ``bullet shape''. The increase of $\Ca$ is also associated with an increase of the bubble relative velocity. Indeed, for $\La = 0.04$ (column (i)), recirculations completely disappear, meaning that the bubble reaches a velocity that is equal to or larger than the maximal flow velocity (here $V = V_\text{b}/J = 2.00$) and that the threshold of the super-fast regime is reached. In contrast, increasing the Laplace number leads to a drop of the relative bubble velocity, as one can see from the reappearance of the flow recirculations at the bubble's front and back in Figs.~\ref{bubble_shape}(b) (ii) and (iii). This decrease in relative velocity is also visible from the increase of the radius of the stagnation circles between Figs.~\ref{bubble_shape}(b) (ii) and (iii). Note that for the highest value of $\La$ ($\La = 378$, Fig.~\ref{bubble_shape}(b)(iii)), an invagination at the rear of the bubble is prominent. As mentioned in the Introduction, the appearance of a reentrant cavity was already observed for bubbles transported in an inertialess flow at large capillary numbers \cite{Chi, Cherukumudi, Martinez_Udell1, Martinez_Udell2, Olbricht, Goldsmith, Nath}. These studies agreed that invaginations are observed for a capillary number larger than $\Ca_{\text{i}}\sim 1$. However, according to the study of Giavedoni \& Saita \cite{Giavedoni}, $\Ca_\text{i}$ drops when increasing the Reynolds number of the flow ($\Ca_{\text{i}} \approx 0.6$ for $\Re = 10$; $\Ca_{\text{i}} \approx 0.25$ for $\Re = 50$). These results agree with the bubble shapes shown in Fig.~\ref{bubble_shape} where invaginations are observed for $\Ca  =1$ with $\La = 0.04$ (column (i)), for $\Ca \geq 0.7$ with $\La = 23$ (column (ii)), and for $\Ca\geq0.4$ with $\La = 378$ (column (iii)).

Increasing further the capillary number leads to even larger bubble deformations and larger relative velocities (see lines (c) and (d) in Fig.~\ref{bubble_shape}). 
As visible from the absence of recirculation, for $\La = 0.04$ and $\Ca\geq 0.4$ (Fig.~\ref{bubble_shape}(b-d)(i)), the super-fast regime is reached. For $\La = 23$ and $\Ca = 0.7$ (Fig.~\ref{bubble_shape}(c)(ii)), even though the front recirculation circles have a relatively small radius, the bubble is still slower than the maximal flow velocity ($V = 1.98$). This is also the case for $\La = 378$ and $\Ca = 0.7$ (Fig.~\ref{bubble_shape}(c)(iii)).  For $\Ca = 1$ and $\La = 23$ (Fig.~\ref{bubble_shape}(d)(ii))), the recirculation at the front disappears, meaning that the super-fast regime is reached. Note that compared to the case where inertia is negligible, i.e., $\La = 0.04$, here, even in the super-fast regime, stationary vortex pairs are visible behind the bubble. However, these vortexes have finite lengths. The presence of recirculations uniquely at the bubble back in the super-fast regime is not directly related to the Laplace number or the inertia of the flow, but appears as soon as there is an invagination at the back of the bubble (data not shown). For the largest value of $\La$ (Fig.~\ref{bubble_shape}(iii)), no super-fast regime could be observed, which means that $V<2$ for all the explored values of $\Ca$. Note that, for $\La = 378$ and large capillary number ($\Ca\gtrsim1$), bubble break-up was observed and no experimental measurements could be performed for $\Ca \sim1$ and $d>0.6$.


\subsection{Bubble velocity}

\begin{figure}[t!]
 \centering
   \includegraphics[width=1\linewidth]{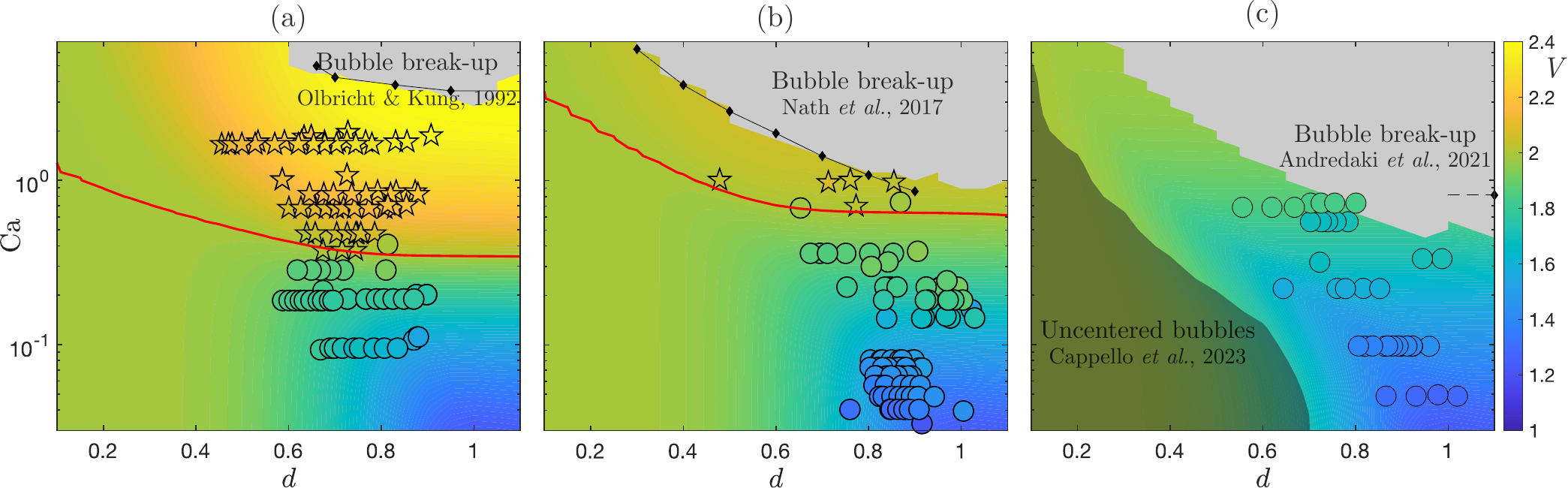}%
 \caption{\textbf{Velocity of centred bubbles}. The dependence of the dimensionless velocity of the bubble with $\Ca$ and $d$ is shown for $\La = 0.04$ \textbf{(a)}, $\La = 23$  \textbf{(b)}, and $\La = 378$ \textbf{(c)}. Numerically derived velocities (background) are compared to experimental measurements (circles and stars). The same colour code is used for the experiments and for the numerical simulations. Star symbols correspond to experiments in which $V\geq2$, circles are used otherwise. The red solid line shows the critical $\Ca^*(d)$ at which $V=2$, as obtained from the numerical simulations. The grey region on the top right corners of the colour maps are regions not reachable with the numerical simulations. The capillary numbers above which bubble break-up is observed, determined by Olbricht \& Kung \cite{Olbricht} at low Reynolds number, by Nath \textit{et al.} \cite{Nath} at $\Re\approx1$, and by Andredaki \textit{et al.} \cite{andredaki2021} at large Reynolds number have been superimposed on (a), (b), and (c), respectively. The shaded area in (c) corresponds to the parameter space for which bubble are no longer stable at the centre-line of the channel.}
 \label{bubble_velocity}
 \end{figure}

The velocities of centered bubbles of varying dimensionless diameter, $d = d_\text{d}/d_\text{h}$, are reported in Fig.~\ref{bubble_velocity}. Experimental measurements (colored dots and stars) are compared with numerical results (colored background). For the sake of visibility, we reported with the red line the threshold for the super-fast regime (i.e. $V=2$), obtained numerically, and we used star symbols for experimental measurements where $V \geq 2$, and circle symbols for $V<2$. An excellent agreement between experiments and numerical simulations is observed, corroborating the accuracy of the experimental measurements and the validity of the numerical approach.
The gray zones correspond to unachievable regimes. Numerically, these zones correspond to sets of parameters for which the radius of curvature at the edges of the rear part of the bubbles decreases below the mesh element size. In practice, this zone corresponds to bubble break-up through a reentrant jet of liquid into the bubble, resulting in bubble break-up. This phenomenon has been observed in the inertia-less flow of bubbles in cylindrical pipes by Olbricht \& Kung \cite{Olbricht}, as well as for non-negligible, yet small, Reynolds numbers ($\Re <  1$) by Nath \textit{et al.} \cite{Nath} and at large Reynolds numbers ($\Re\approx500$) by Andredaki \textit{et al.} \cite{andredaki2021} only for large bubbles ($d\gg1$). 
The dependence of the bubble break threshold on the capillary number $\Ca$ and the bubble confinement $d$ obtained by the studies mentioned above have been added to Fig.~\ref{bubble_velocity} (black diamonds). 
As the gray regions of Figs.~\ref{bubble_velocity}(a)-(c) match fairly the bubble-break-up zones, it is reasonable to assume that the numerical calculation limits coincide with the bubble-break-up thresholds for each map.

Interestingly, for the two smallest values of $\La$ (i.e., $\La = 0.04$ and $\La = 23$), the critical capillary number, $\Ca^*$, above which the super-fast regime is observed, decreases when increasing $d$ before reaching a plateau at large $d$. Comparing the results obtained for the three values of $\La$, we observe that, in agreement with what was previously observed, the velocity of the bubble decreases when $\La$ increases for any fixed bubble size $d$. Hence, for $\La = 23$, $\Ca^*$ is larger than for $\La = 0.04$. Furthermore, for the highest value of $\La$ (see Fig.~\ref{bubble_velocity}(c)), no super-fast regime could be observed, neither numerically nor experimentally.

Note that, for the case $\La = 378$, the shaded area of Fig~\ref{bubble_velocity}(c) corresponds to a zone in which the center position of the bubbles is no longer stable because of inertial migration \cite{cappello2023beads} that pushes the bubble towards an off-centered equilibrium position \cite{segre1962behaviour}. This situation is not observed for $\La = 0.04$ and $\La = 23$ for which the centered bubbles are always stable because the deformation-induced migration force, oriented toward the channel center line, dominates over the inertial migration force.


\subsection{Modelling the velocity of long bubbles}

Although the article focuses on small bubbles (i.e. $d<1$), investigating the situation of a long bubble ($d\gg1$) gives insight to understand the dependence of the bubble velocity with the capillary and Laplace numbers, as well as to understand the mechanisms underlying the existence of the super-fast regime. Note that large bubbles are always centered in the channel.

Following the seminal study of Lac \& Sherwood \cite{Lac}, we consider a laminar annular flow of two fluids driven by a pressure gradient $\delta_x p$ in a capillary of diameter $d_\text{h}$. This coflow configuration corresponds to the limit of an infinitely long bubble, that is for $d\rightarrow \infty$. The inner fluids, of viscosity $\lambda \mu$, occupy a cylinder of radius $r_\text{b}<d_\text{h}/2$, and the outer liquid, of viscosity $\mu$, occupies the annular film of thickness $\delta = d_\text{h}/2-r_\text{b}$. Continuity of the longitudinal fluid velocity ${u}$ and of the tangential stress yields: 
\begin{equation}
u(r) = \left\{
    \begin{array}{ll}
        \frac{\delta_x p}{4 \mu}(r^2-\left(\frac{d_\text{h}}{2}\right)^2) \quad & d_\text{h}/2 -\delta \leq r\leq d_\text{h}/2, \\
        \frac{\delta_x p}{4 \lambda \mu}\left[r^2+(\lambda-1)\left(\frac{d_\text{h}}{2}-\delta \right)-\lambda\left(\frac{d_\text{h}}{2}\right)^2\right]\quad & 0 \leq r\leq d_\text{h}/2 -\delta.
    \end{array}
\right.
\end{equation}

The superficial velocity is
\begin{equation}
J = \frac{8}{d_\text{h}^2}\int^{d_\text{h}/2}_0 u(r) r \text{d}r,
\end{equation}
and if the inner fluid mean velocity is
\begin{equation}
V_\text{b} = \frac{2}{\left(\frac{d_\text{h}}{2}-\delta \right)^2}\int^{d_\text{h}/2-\delta}_0 u(r) r \text{d}r.
\end{equation}

Hence the dimensionless velocity of the inner fluid reads : 
\begin{equation}
V_{d\rightarrow \infty} = \frac{V_\text{b}}{J} = \frac{2\lambda + (1-2\lambda)\left[1-2\delta/d_\text{h}\right]^2}{\lambda+(1-\lambda)[1-2\delta/d_\text{h}]^4}.
\label{Vinner}
\end{equation}

According to Eq.~(\ref{Vinner}), the dimensionless velocity of the inner fluid can be larger than 2 if $\lambda<1/2$ \cite{Lac}.
For the bubbles considered in this study, that is, $\lambda\ll1$ in experiments and $\lambda = 0$ in numerical simulations, Eq.~(\ref{Vinner}) simplifies to : 

\begin{equation}
V_{d\rightarrow \infty} = \left[ 1 - \frac{2 \delta}{d_\text{h}}\right]^{-2},
\label{Vbubble}
\end{equation}
which is a monotonically increasing function of $2 \delta/d_\text{h}$.
Hence, capturing the dependence of $2 \delta/d_\text{h}$ with the capillary and Laplace numbers enables the derivation of the velocity of long bubbles. 
According to the seminal study work of Bretherton \cite{Bretherton}: 
\begin{equation}
\frac{2 \delta}{d_\text{h}}\sim \Ca^{2/3}.
\label{eq_Bretherton}
\end{equation}
Yet, this expression has been derived for long bubbles in a Stokes flow and for $\Ca\ll1$. Aussilous \& Qu\'er\'e expanded the Bretherton law (Eq.~(\ref{eq_Bretherton})) to larger values of capillary numbers \cite{Aussillous}: 
\begin{equation}
\frac{2 \delta}{d_\text{h}} = \frac{1.43 \,\Ca^{2/3}}{1+3.35\, \Ca^{2/3}}.
\label{eq_aussillous}
\end{equation}
More recently, the study of Langewish \& Buongiorno \cite{Langewisch} extended further the range of validity of the Bretherton law by taking into account inertia. They proposed the following formula, 
\begin{subequations}
\begin{equation}
   \frac{2 \delta}{d_\text{h}} =  \frac{1.375 \,\Ca^{2/3}}{1+2.86\left[1+\Phi(\Re)\right]\Ca^{0.764}},\quad \quad  
   \end{equation}
   \begin{equation}
\text{with} \quad  \Phi(\Re) = \left[32.05\,\Re^{-0.593}+4.564\times10^{-5}\,\Re^{1.909}\right]^{-1},
\end{equation}
\label{eq_Buongiorno}
\end{subequations}
by fitting the numerical results obtained for $d\gg 1$, $\Ca \le 2$ and $\Re \le 120$.

\begin{figure}[t!]
 \centering
   \includegraphics[width=1\linewidth]{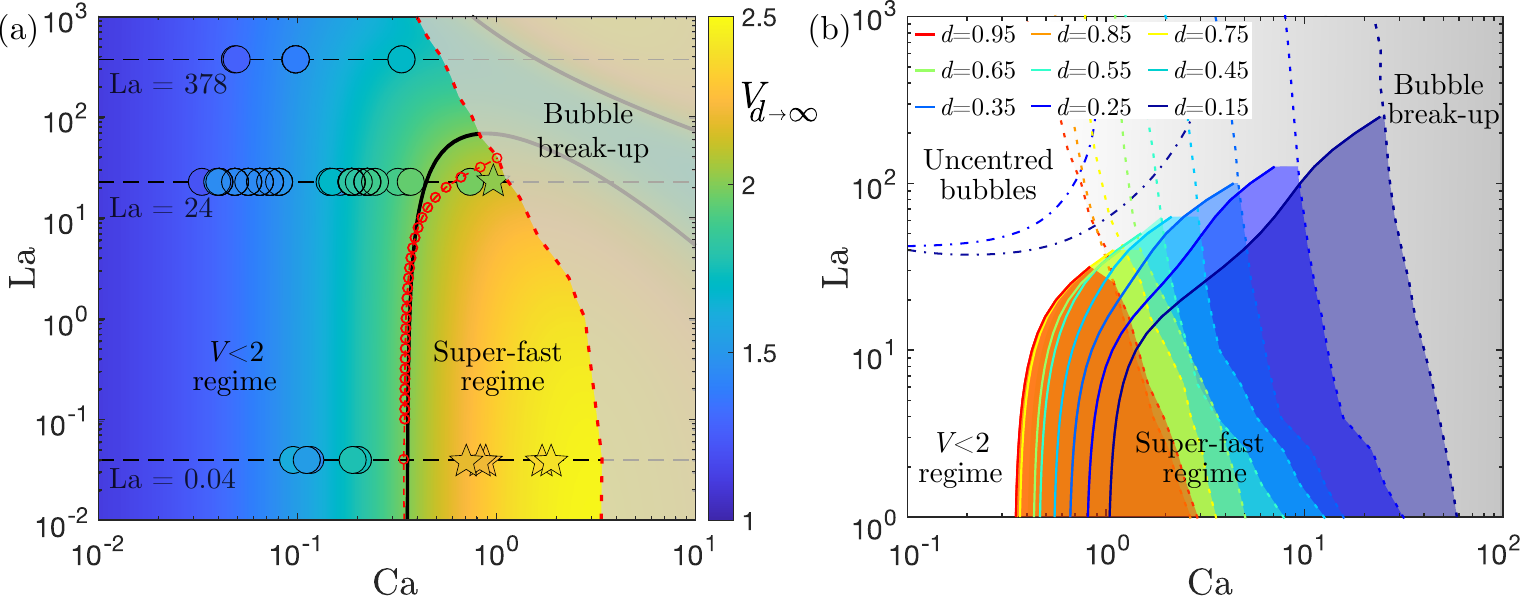}%
 \caption{\textbf{Super-fast regime in the $\La-\Ca$ space.} \textbf{(a)} Color-coded bubble velocity for varying Laplace and capillary numbers as obtained with the analytical coflow model. The blacks solid lines are the contours verifying $V_{d\rightarrow \infty}=2$. Circles and stars are experimental measurements of long bubbles ($d\geq0.9$) with velocities that verify $V<2$ and $V\geq2$, respectively, using the same color code. The critical capillary number $\Ca^*$ above which the bubbles are in the super-fast regime obtained from the numerical simulations for $d=1$ and for varying $\La$, are shown by the red circles. The shaded area corresponds to the region where bubble break-up is expected from the numerical simulations. \textbf{(b)} For varying bubble size $d$, the region in the $\Ca-\La$ space corresponding to the super-fast regime are shown by the coloured areas. These regions are delimited, by the critical capillary number $\Ca^*$ at lower values of $\Ca$ and by bubble break-up $\Ca_b$ at larger values of $\Ca$. The dependence of $\Ca^*$ with $\La$ are shown by the coloured solid lines and the bubble break-up regime is observed at the right of the dashed coloured lines. The two dotted-dashed lines correspond to the limit of stability of the centred position of the bubbles for $d=0.15$ and $d = 0.25$ (i.e. the two worth case situations). The regions located above the curves correspond to unstable centred bubbles while elsewhere the bubbles are stable at the channel centre-line.}
 \label{fig_long_bubble}
 \end{figure}

Combining Eqs.~(\ref{Vbubble}) and (\ref{eq_Buongiorno}) with $\Re = \La\Ca$, we can calculate the long bubble velocity as a function of both $\Ca$ and $\La$, as shown in the color map of Fig.~\ref{fig_long_bubble}(a). The contours $V_{d\rightarrow \infty} = 2$ are drawn in solid black lines, allowing comparison with the experimental measurements obtained for large bubbles ($d\gtrsim 0.9$) as illustrated by circles when $V<2$ or stars when $V\geq2$.

According to the analytical coflow model, the relative velocity of the bubble (colored background in Fig.~\ref{fig_long_bubble}(a)) does not depend on the flow inertia for $\La \lesssim 5$, and therefore the critical capillary number above which the super-fast regime is reached is roughly constant and equal to $\Ca^* = 0.34$. 
For $\La\leq 70$ the relative velocity of the bubbles first increases before decreasing with $\Ca$, and a first branch of the contour $V_{d\rightarrow \infty} = 2$ shows that $\Ca^*$ first increases, then decreases with $\La$ with a maximum at $(\Ca^*,\La)=(0.8,70)$. However, when superimposing the region where bubble break-up occurs according to the absence of solutions in our numerical simulations (shaded area), it is observed that the decrease of the relative velocity for $\La \leq 70$ cannot be observed in practice.
For $\La \geq 70$, the relative velocity increases monotonically with $\Ca$, and a second branch of the contour $V_{d\rightarrow \infty} = 2$ shows that $\Ca^*$ should decrease with $\La$, even though it entirely lies in the shaded area corresponding to bubble break-up and cannot be observed in practice.

Figure~\ref{fig_long_bubble}(a) finally shows the critical capillary number $\Ca^*$ obtained with the numerical simulations for large bubbles with $d=1$ (red circles). We observe that the analytical flow model is reliable up to $\La \approx 10$, above which it significantly overestimates $\Ca^*$, essentially because it ignores the role of bubble caps and associated recirculation flows in slowing the velocity of the bubble.
In summary, the super-fast regime can only be observed in a very specific area of the $\La-\Ca$ space delimited by the red circles for $\Ca^*$ and by the dashed red line for $\Ca_\text{b}$ corresponding to the threshold for the bubble break-up. Noticeably, $\Ca^*$ and $\Ca_\text{b}$ cross each other at a critical Laplace number $\La_{\rm c} \approx 40$ above which no super-fast regime cannot be observed for large bubbles. 

\subsection{Super-fast regimes for small bubbles}

Figure~\ref{fig_long_bubble}(b) builds on the existence of the super-fast regime (colored areas) in the $\La-\Ca$ space for varying bubble diameters ($d$ decreases from 0.95 in red to 0.15 in blue). Again, these regions are delimited by the $\Ca^*$ and the $\Ca_\text{b}$ curves (solid and dashed lines, respectively). We note an increase of the super-fast regime area when decreasing $d$ as, for all Laplace numbers, the increase of $\Ca_\text{b}$ with $d$ is larger than the increase of $\Ca^*$. Consequently, the critical Laplace number $\La_{\rm c}$ increases as $d$ decreases. For the lowest value of $d=0.15$, the highest value of $\La_{\rm c}\approx 250$, which explains why no super-fast regime was observed for $\La = 378$ in our experiments.

We also added the stability curves of the center positions for the two smallest values of $d$ (i.e, $d = 0.15$ and $d=0.25$). The regions located on the top left of the curves correspond to unstable centered bubbles, whereas elsewhere the bubbles are stable at the channel center line. Since there is no overlap of these curves with the colored regions, we can state that the super-fast bubble regimes always lie in the region for which their centered position is stable. 

\section{Conclusion}

\textcolor{black}{In this study, we focused on the velocity of bubbles transported by pressure-driven flow in cylindrical microchannels. By combining experimental and numerical approaches, we provided the first experimental proof of the so-called “super-fast” regime, where bubbles achieve velocities exceeding the maximum flow velocity and we  thoroughly rationalized the conditions required to observe this regime.}

\textcolor{black}{We demonstrate that it is constrained to a specific range of capillary ($\Ca$)  and Laplace ($\La = \Re/\Ca$) numbers, which highlight the critical interplay between deformability and inertia in bubble dynamics. 
The super-fast regime is bounded by two critical capillary numbers: a lower threshold ($\Ca^*$) marking the onset of the regime and an upper threshold ($\Ca_\text{b}$) beyond which bubble breakup occurs due to tip streaming at the bubble's rear where an invagination is observed for $\Ca>\Ca_\text{i}>\Ca_\text{b}$. These two limits define the operational window where bubbles remain stable and reach maximum velocities. We further demonstrated that $\Ca^*$ and $\Ca_\text{b}$ are functions of $\La$, crossing at a critical value $\La_\text{c}$ above which the super-fast regime does not exist anymore, indicating that large inertia inhibits the observation of the super-fast regime.}

\textcolor{black}{Additionally, we proposed an analytical model that predicts the velocity of large bubbles ($d \to \infty$) and the evolution of $\Ca^*$ for these bubbles. This model gives insights on the origins of the super-fast regime that results from the interplay between two aspects: First, as bubbles are less viscous than the continuous phase, the combined bubble-annulus section moves faster than a section solely composed of the continuous phase for the same pressure drop. The velocity of the bubble is therefore always larger than the mean velocity of the continuous phase. Second, as the capillary number increases, the bubbles gets more and more elongated and the gap between the bubble and the wall widens, moving the bubble closer to the highest-velocity region of the channel and leading to an additional increase of its velocity. For sufficiently large gaps (\textit{i.e}, large capillary numbers) the resulting increase of the bubble velocity is large enough to reach the super-fast regime.}

\textcolor{black}{Lastly, we expand our analysis to small bubble $d\leq1$, and presented a complete phase diagram of the super-fast regime in the $\Ca-\La$ space for varying bubble confinement ($d$). We showed that decreasing $d$ expands the area of the region where the super-fast regime is observed, increases the value of $\La_\text{c}$, but also raises $\Ca^*$, making experimental observation more challenging.}

\textcolor{black}{Extending these findings to practical applications, such as multiphase microreactors or mass transfer systems, this research provides a foundation for optimizing processes involving gas-liquid flows. Future work could explore additional factors, such as surfactants or complex channel geometries, to broaden the applicability of these insights.}

\paragraph{Acknowledgements.}
JC acknowledge support by F.R.S.-FNRS under the research grants CR n◦40017301 “G.El.In.Flow". This project has received funding from 747 Fédération Wallonie-Bruxelles (ARC ESCAPE project). We thank the Micro-milli service platform (ULB) for the access to their experimental facilities.

\paragraph{Competing Interests.}
There is no competing interest to declare.

\paragraph{Authors contributions.} J.C., J.R. and B.S. designed research; J.C. and J.R. performed research; J.C. and J.R. analyzed data; J.C. performed the experiments; J.R. and B.S. developed the numerical model; all authors wrote the manuscript.
\paragraph{Data and code availability.}
The raw data and the code supporting the findings of this study are available from the corresponding author upon reasonable request.

\bibliographystyle{unsrt}
\bibliography{biblio_bbbubble.bib}

\end{document}